\documentclass[fleqn,usenatbib]{mnras}

\usepackage{newtxtext,newtxmath}

\usepackage[T1]{fontenc}
\usepackage{ae,aecompl}

\DeclareRobustCommand{\VAN}[3]{#2}
\let\VANthebibliography\thebibliography
\def\thebibliography{\DeclareRobustCommand{\VAN}[3]{##3}\VANthebibliography}

\usepackage{graphicx}	
\usepackage{amsmath}	
\usepackage{natbib}
\usepackage[normalem]{ulem} 

\newcommand{\bs}[1]{\boldsymbol{#1}}
\newcommand{\bcdot}{\boldsymbol{\cdot}}
\newcommand{\bnabla}{\boldsymbol{\nabla}}
\newcommand{\btimes}{\boldsymbol{\times}}
\newcommand{\dps}{\displaystyle}

\title[Cosmic ray hydrodynamics closure relations]{Comparing different closure relations for cosmic ray hydrodynamics}

\author[T.\ Thomas and C.\ Pfrommer]{
T. Thomas,$^{1,2}$\thanks{E-mail: tthomas@aip.de (TT)}
C. Pfrommer,$^{1}$
\\
$^{1}$ Leibniz-Institute for Astrophysics Potsdam (AIP), An der Sternwarte 16, 14482 Potsdam, Germany\\
$^{2}$ Institute of Physics and Astronomy, University of Potsdam, Karl-Liebknecht-Str. 24/25, 14476 Golm, Germany
}

\date{Accepted XXX. Received YYY; in original form ZZZ}

\pubyear{2020}

\begin{document}
\label{firstpage}
\pagerange{\pageref{firstpage}--\pageref{lastpage}}
\maketitle

\begin{abstract}
Cosmic ray (CR) hydrodynamics is a (re-)emerging field of high interest due to the importance of CRs for the dynamical evolution of the interstellar, the circumgalactic, and the intracluster medium. In these environments, CRs with GeV energies can influence large-scale dynamics by regulating star formation, driving galactic winds or by altering the pressure balance of galactic halos. Recent efforts have moved the focus of the community from a one-moment description of CR transport towards a two-moment model as this allows for a more accurate description of the microphysics impacting the CR population. Like all hydrodynamical theories, these two-moment methods require a closure relation for a consistent and closed set of evolution equations. The goal of this paper is to quantify the impact of different closure relations on the resulting solutions. To this end, we review the common P1 and M1 closure relations, derive a new four-moment H1 description for CR transport and describe how to incorporate CR scattering by Alfv\'en waves into these three hydrodynamical models. While there are significant differences in the transport properties of radiation in the P1 and M1 approximations in comparison to more accurate radiative transfer simulations using the discrete ordinates approximation, we only find small differences between the three hydrodynamical CR transport models in the free streaming limit when we neglect CR scattering. Most importantly, for realistic applications in the interstellar, circumgalactic or intracluster medium where CR scattering is frequent, these differences vanish and all presented hydrodynamical models produce the same results.
\end{abstract}



\section{Introduction}
\label{sec:introduction}

CRs are a divers population of energetic particles with energies ranging from MeV to PeV \citep{2013Zweibel}. While low-energy CRs are responsible for (some of) the ionisation of gas in molecular clouds \citep{2009Padovani} and high energy CRs with energies $\gtrsim$ TeV provide a window into the dynamics of CRs through observations with imaging air/water Cherenkov telescopes \citep{2007Strong}, the bulk energy-carrying GeV CRs influence the kinematics and thermodynamics of various astrophysical environments \citep{2017Zweibel}. CRs injected into the interstellar medium at the shocks of supernovae can leave their injection site, lift gas out of a galactic disk, and drive mass-loaded galactic scale winds \citep{1991Breitschwerdt, 2012Uhlig, 2016Recchia, 2020Dashyan, 2021Rathjen}. Because this gas is removed momentarily or completely out of the galaxy it cannot participate in subsequent star formation processes \citep{2016Girichidis, 2016Simpson, 2017Pfrommer, 2018Farber, 2021Semenov}. Furthermore, the ejected gas in combination with the CRs provides an additional pressure in the circumgalactic medium, which alters the accretion of fresh gas out of the intergalactic medium \citep{2020Buck}. To correctly model these astrophysical processes, it is important to describe the physics of CRs correctly. This is a difficult task because CRs only rarely interact with their surroundings through particle-particle collisions while they primarily interact indirectly with the gas through scatterings at electromagnetic fields \citep{2017Zweibel}. If the surrounding gas can be described in terms of magneto-hydrodynamics (MHD), CRs provide a pressure that is directed perpendicular to the MHD magnetic field \citep{2013Zweibel}. Parallel to this magnetic field, CR transport is governed by their interactions with small-scale perturbations with typical length scales comparable to the gyroradius of CRs. On these length scales MHD is not applicable and kinetic plasma physics dictates the evolution of CRs \citep{BookKulsrud}. Various plasma processes such as the gyroresonant \citep{1969Kulsrud}, the intermediate scale \citep{2021Shalaby}, or Bell's instability \citep{2004Bell} can strongly influence the propagation of CRs.

To model the transport of CRs on scales significantly larger than their gyroradius, we need to adopt drastic approximations. These require the development of theoretical models and numerically tractable algorithms that are efficient enough so that the resulting model is able to bridge the separation in length scales from small-scale micro physics to the large-scale astrophysical environment. These requirements are rather challenging to achieve with a kinetic description but can be readily provided by hydrodynamical approximations. The clear downside of hydrodynamical models is that they coarse-grain kinetic physics, thus possibly losing some of the important plasma-kinetic effects. Yet, the emergence of large-scale simulations employing CR hydrodynamics \citep{2016Girichidis, 2017Pfrommer, 2019Chan} and their ability to approximately model CR microphysics  \citep{2019Thomas, 2020Hopkins} rather than completely neglecting CRs at all proves the success of CR hydrodynamics. 

The first CR hydrodynamical theories were one-moment hydrodynamical models in which one scalar quantity (either the CR energy or number density) is used to describe the entire CR population and its propagation properties \citep{1991Breitschwerdt}. The resulting transport phenomena can be categorised into (i) \textit{CR advection}, which describes CR confinement by a large scale magnetic field to the gas so that CRs follow the bulk movement of the gas, (ii) \textit{CR streaming}, which assumes frequent CRs scatterings with small-scale electromagnetic waves so that CRs are confined to move with these waves along a magnetic field line, and (iii) \textit{CR diffusion}, which applies to infrequent CR scatterings that are not able to maintain bulk kinetic motions of CRs but strong enough so that the remaining transport can be expressed through a diffusion process \citep{2020Thomas}. 
While the theoretical description of all these processes is well established, only CR advection and diffusion can be numerically described by standard methods with small modifications to account, e.g., for the anisotropic nature of CR transport \citep{2003Hanasz, 2011Sharma, 2012Yang, 2014Girichidis, 2016Pakmor, 2018Butsky, 2019Dubois}. Finding a stable and accurate numerical discretisation of CR streaming in the one-moment picture proves to be difficult \citep{2009Sharma}. This prompted the (re-)development of two-moment hydrodynamical models, which also include the flux of the scalar quantity as an independent and evolved quantity \citep{1985Webb,2018Jiang,2019Thomas}. These second-order models solve the problem of how to numerically discretise the CR streaming effect but come with the increased demand for effective theoretical descriptions of the CR microphysics to get a self-consistent evolution of the CR flux. Additionally, they provide convenient tools that can describe CR diffusion more accurately and do not completely break down if CR scattering is less frequent. 

Like all hydrodynamical models, these two-moment models suffer from a closure problem: an $n$-th moment hydrodynamical model depends on the $(n+1)$-th moment or, in other words, every hydrodynamical model requires a closure for the next-higher order moment in terms of the moments it is evolving. Thus, two-moment models, which evolve only the zeroth and first moment, require information about the second moment of the CR distribution for a consistent set of closed evolution equations. This problem was implicitly circumvented in the theories of \citet{2018Jiang} and \citet{2019Thomas} by using a Taylor (or Legendre) expansion of the CR distribution up to first order while assuming that any higher-order expansion coefficients vanish. This enables them to calculate the required second moment in terms of the zeroth and first moment directly without additional information or physical insight. This procedure is also known in the theory of radiation transport and called the P1 approximation. In this context the P1 approximation has been shown to exhibit some shortcomings: the P1 approximation implicitly assumes that scattering is frequent and removes any higher-order expansion coefficients. This leads to some problematic properties of the P1 approximation: (i) while the maximum propagation speed of radiation is the speed of light $c$, it is artificially reduced to $c / \sqrt{3}$ in the P1 approximation \citep{2000Olson} and (ii) if scattering is infrequent, contrary to the assumption, the P1 approximation can produce negative number or energy densities \citep{1976Kershaw}. Both shortcomings can be avoided by using an M1 approximation which employs other means to calculate the second moment \citep{1984Levermore}. Recently, this idea originally developed for radiation hydrodynamics was translated to the case of CR transport \citep{2021Hopkins}.

In this paper, we compare the P1 and M1 approximations for CR transport in terms of the resulting evolution for a given CR or radiation distribution. We quantity the differences with the goal to judge which of the two approximations yields the most accurate solution. To this end, we discuss and review the derivations of the P1 and M1 approximations for CRs and radiation in Section~\ref{sec:hydrodynamics}. We additionally develop a new four-moment approximation named the H1 approximation for CR-transport and present simplified models of CR scatterings with Alfv\'en waves for all three hydrodynamic models. In Section~\ref{sec:numerics}, we illustrate the differences of the closures with a set of numerical simulations where we use the same simulation setup but alter the used transportation model to evolve radiation, CR hydrodynamics without and with scatterings. We close this paper with a discussion in Section~\ref{sec:discussion}. The goal-oriented reader may want to directly jump to Section~\ref{sec:numerics} for the main results of this study. We use $\mathbfit{a} \mathbfit{b}$ to denote the dyadic product of the two vectors $\mathbfit{a}$ and $\mathbfit{b}$.

\section{CR and Radiation Hydrodynamics}
\label{sec:hydrodynamics}
In this section, we present the theoretical foundations required to understand our numerical experiments. We review the derivations of the two-moment P1 and M1 hydrodynamical closure relations and derive a new four-moment H1 approximation for the case of CR transport. This will enable us to judge whether it is worth to further explore this route and to derive even higher-order closure relations to capture more detailed dynamics of CRs. 

In our discussion, we focus on a mono-energetic ultra-relativistic population of CRs. In this limit, the Boltzmann transport equation for a distribution of particles can be written as:
\begin{equation}
    \frac{1}{c}\frac{\partial f}{\partial t} + \mathbfit{d} \bcdot \bnabla f =  \left.\frac{1}{c}\frac{\partial f}{\partial t} \right\vert_\mathrm{scatt} \label{eq:boltzmann},
\end{equation}
where $c$ is the speed of light and $\mathbfit{d} = \bs{\varv} / c$ is the direction of the CR velocity $\bs{\varv}$. We place all terms describing the interaction of CRs with electro-magnetic fields on the right-hand side and use this term as a placeholder, which will be specified in Section~\ref{sec:scattering}. In this notation, the Boltzmann equation is formally equivalent to the radiative transfer equation when $f$ is interpreted as the specific intensity of radiation and $\mathbfit{d}$ is the propagation direction of radiation. This degeneracy is broken once we account for CRs gyrating around the magnetic field. Assuming that the population of CRs is equally distributed in terms of the gyration phase angle, Eq.~\eqref{eq:boltzmann} can be simplified to the \textit{focused} transport equation \citep{BookZank}:
\begin{equation}
    \frac{1}{c}\frac{\partial f}{ \partial t} + \mu \mathbfit{b} \bcdot \bnabla f + (\bnabla \bcdot \mathbfit{b}) \frac{1 - \mu^2}{2} \frac{\partial f}{\partial \mu} = \left.\frac{1}{c}\frac{\partial f}{\partial t} \right\vert_\mathrm{scatt} \label{eq:vlasov},
\end{equation}
where $\mathbfit{b}=\mathbfit{B}/B$ is the direction of the magnetic field, $B=\sqrt{\mathbfit{B}^2}$ is the magnitude of the magnetic field, and $\mu = \mathbfit{b} \bcdot \mathbfit{d}$ is the pitch angle cosine of the CRs so that the mono-energetic, gyrotropic CR distribution attains the following dependence: $f=f(\mathbfit{x},\mu)$, where $\mathbfit{x}$ is the space coordinate. The second term states that this equation solely describes transport of CRs through space along magnetic field lines. The third term is the magnetic focusing term and describes the change of a particle's pitch angle caused by a spatially varying magnetic field through the $\bnabla \bcdot \mathbfit{b}$ factor. This term has a simple physical interpretation: if $\bnabla \bcdot \mathbfit{b} > 0$ then magnetic field lines diverge and CRs are beamed in the direction of the magnetic field. In the opposite situation, when magnetic field lines converge or $\bnabla \bcdot \mathbfit{b} < 0$, CRs are beamed in the opposite direction of the magnetic vector field. Another useful way to write this equation is given by:
\begin{equation}
    \frac{1}{c}\frac{\partial f}{\partial t} +  (\mathbfit{b} \bnabla) \bs{:} \left[  \mu f  \mathbf{1} + \frac{1 - \mu^2}{2} \frac{\partial f}{\partial \mu} (\mathbfit{b} \mathbfit{b} - \mathbf{1}) \right] = \left.\frac{1}{c}\frac{\partial f}{\partial t} \right\vert_\mathrm{scatt}.
    \label{eq:vlasov2}
\end{equation}
This result can be derived by realising that $\mathbfit{b}\bcdot\mathbfit{b} = 1$ and $(\mathbfit{b} \mathbfit{b}) \bs{:} \bnabla \mathbfit{b} = \mathbfit{b}\bcdot \bnabla (\mathbfit{b}\bcdot\mathbfit{b})/2 =0$. The advantage of this formulation is that it only contains one spatial derivative and eliminates the $\bnabla \bcdot \mathbfit{b}$ term which is numerically challenging to model.

Here, the starting point of the derivation of CR transport differs from the one presented in \citet{2019Thomas} by the absence of any pseudo forces. These pseudo forces enter into the focused transport equation if it is formulated in a non-inertial frame such as the comoving frame of an MHD fluid. Instead, in this work we solely focus on the anisotropic transport of CRs along magnetic fields. Consequently, we omit those pseudo-force terms for reasons of transparency of the derivation and note that they can be added without changing any of our conclusions. 

\subsection{Two-moment hydrodynamics and the closure problem}

We first derive a two-moment hydrodynamical approximation for CR transport as described by Eq.~\eqref{eq:vlasov} in a general setting. Such a theory aims to find evolution equations for the first two moments of $f$, which are defined by
\begin{align}
    f_0 &= \frac{1}{2} \int_{-1}^{1} \mathrm{d}\mu~ f, \label{eq:f_0} \\
    f_1 &= \frac{1}{2} \int_{-1}^{1} \mathrm{d}\mu~ \mu f \label{eq:f_1},
\end{align}
in such a way that their evolution depends on each other while $f_0$ and $f_1$ remain independent quantities. In such a two-moment approximation, the information about the $\mu$-structure of $f$ is encoded in the moments $f_0$ and $f_1$, which in general implies a loss of information because not all the potential structure of $f$ can be mapped onto the first two moments. However, if the higher-order moments (representing a larger degree of anisotropy in phase space) are maintained at a small level by some physical (scattering) process, it is sufficient to evolve $f_0$ and $f_1$ and to capture the essentials of the physics.

The individual evolution equations can be extracted from Eq.~\eqref{eq:vlasov} by multiplying both sides of Eq.~\eqref{eq:vlasov} with $1$ and $\mu$, respectively, and averaging the result over $\mu$ space to arrive at
\begin{align}
    \frac{1}{c}\frac{\partial f_0}{ \partial t} &+ \bnabla \bcdot (\mathbfit{b} f_1) = \left.\frac{1}{c}\frac{\partial f_0}{\partial t} \right\vert_\mathrm{scatt}, \label{eq:two_moment_1}\\
    \frac{1}{c}\frac{\partial f_1}{\partial t} &+ \mathbfit{b} \bcdot \bnabla f_2 + (\bnabla \bcdot \mathbfit{b}) \frac{3 f_2 - f_0}{2} = \left.\frac{1}{c}\frac{\partial f_1}{\partial t} \right\vert_\mathrm{scatt},
    \label{eq:two_moment_2}
\end{align}
where we define the second moment $f_2$ by
\begin{equation}
    f_2 = \frac{1}{2} \int_{-1}^{1} \mathrm{d}\mu~ \mu^2 f(\mu). \label{eq:f_2}
\end{equation}
With these equations at hand, the first moment $f_1$ can also be interpreted as the flux of $f_0$. The evolution of $f_0$ depends on $f_1$ according to Eq.~\eqref{eq:two_moment_1} and the evolution of $f_1$ depends itself on $f_2$ (Eq.~\ref{eq:two_moment_2}), which is an unknown quantity. Consequently, Eqs.~\eqref{eq:two_moment_1} and \eqref{eq:two_moment_2} do not form a closed set of equations and cannot be solved without additional information on $f_2$.

This is the so called \textit{closure problem} that has the obvious solution to choose a \textit{closure relation} for $f_2$ in form of a closed expression that depends on $f_0$ and $f_1$. We now present the P1 and M1 closure relations for CR two-moment hydrodynamics which originate from radiation hydrodynamics. 

\subsection{The P1 approximation}
The P1 approximation is the simplest hydrodynamical approximation of Eq.~\eqref{eq:vlasov} and results in a simple closure relation for the second moment. This approximation adopts a linear function in pitch-angle for the distribution $f$. We use the notation of \citet{2019Thomas} and define
\begin{align}
f_\text{P1} = f_0 + 3 \mu f_1, \label{eq:P1}
\end{align}
which complies with the definitions of the first two moments in Eqs.~\eqref{eq:f_0} and \eqref{eq:f_1}.
All other moments in the Taylor expansion of $f$ are assumed to be vanishingly small so that their contribution to the resulting dynamics of $f$ can be neglected. The second moment in this approximation as calculated by the integral in Eq.~\eqref{eq:f_2} using $f_\text{P1}$ from Eq.~\eqref{eq:P1} is given by:
\begin{align}
    f_2 = \frac{1}{3} f_0,
\end{align}
which forms the closure relation for the P1-approximation. The evolution equations in Eqs.~\eqref{eq:two_moment_1} and \eqref{eq:two_moment_2} simplify to:
\begin{align}
    \frac{1}{c}\frac{\partial f_0}{ \partial t} &+ \bnabla \bcdot (\mathbfit{b} f_1) = \left.\frac{1}{c}\frac{\partial f_0}{\partial t} \right\vert_\mathrm{scatt}, \label{eq:P1_1}\\
    \frac{1}{c}\frac{\partial f_1}{\partial t} &+ \frac{1}{3}\mathbfit{b} \bcdot \bnabla f_0 = \left.\frac{1}{c}\frac{\partial f_1}{\partial t} \right\vert_\mathrm{scatt}.\label{eq:P1_2}
\end{align}
The second term in Eq.~\eqref{eq:P1_2}, $\mathbfit{b} \bcdot \bnabla f_0/ 3$, can be interpreted as a pressure term for an isothermal fluid with sound speed $c / \sqrt{3}$. In a constant magnetic field these equations are a set of linear hyperbolic differential equations whose characteristics travel with velocities $\pm c / \sqrt{3}$. Building upon this insight, we can construct an alternative distribution function that yields the same moments as $f_\mathrm{P1}$ in Eq.~\eqref{eq:P1} and an identical evolution of $f_0$ and $f_1$ as described by Eqs.~\eqref{eq:P1_1} and \eqref{eq:P1_2} but highlights the two-stream property:
\begin{align}
f_\text{2-stream} 
=\left(f_0-\sqrt{3}f_1\right) \delta \left(\mu + \frac{1}{\sqrt{3}}\right) + \left(f_0+\sqrt{3}f_1\right) \delta \left(\mu - \frac{1}{\sqrt{3}}\right), \hspace{-4pt} \label{eq:delta_P1}
\end{align}
where $\delta$ denotes Dirac's $\delta$ distribution. Thus, the P1 approximation coincides with a two-stream approximation.
Consequently, it can not capture beam-like CR transport where all CRs are travelling with velocities $\bs{\varv}$ projected onto magnetic field that have $\vert \bs{\varv} \cdot \mathbfit{b} \vert \sim c$ or $\vert \mu \vert \sim 1$ on average. A further shortcoming of the P1 approximation is that information is only propagated at $\pm c / \sqrt{3}$ but not at any smaller or higher velocity. Both points are serious problems of the P1 approximation as those propagation modes are included in Eq.~\eqref{eq:vlasov} where the fastest velocity is equal to $\pm c$ but information is also propagated at any velocity in between $-c$ and $+c$. The reason for this shortcoming is the linear approximation in Eq.~\eqref{eq:P1}: the distribution $f$ must be peaked toward $\mu \sim \pm 1$ in order to realise a beam transport mode. This anisotropic pitch-angle distribution cannot be captured by any combination of $f_0$ and $f_1$. However, the clear advantage of the P1 approximation is its simplicity and correctness when CR scattering as described by $\partial f/\partial t \vert_\mathrm{scatt}$
is strong and damps all higher-order moments in the pitch-angle distribution.

The distribution $f$ correspond to a probability density and has to be non-negative if it describes a physical population of particles. A natural question to ask is whether we can find such a non-negative distribution for a given set of moments.
\citet{1976Kershaw} derived general conditions on the first three moments that---provided they are fulfilled---guarantee 
the existence of such a non-negative distribution $f$. In our notation, the first three conditions are
\begin{align}
    0 &\leq f_0, \label{eq:kershaw_0} \\
 \vert f_1 \vert &\leq  f_0, \label{eq:kershaw_1} \\
 \left(\frac{f_1}{f_0}\right)^2 &\leq \frac{f_2}{f_0} \leq 1. \label{eq:kershaw_2}
\end{align}
These conditions provide a mathematical explanation for the failure of the P1 approximation: in this model, we have $f_2 = f_0/3$ and the third condition is violated if $\vert f_1 \vert > f_0 / \sqrt{3}$. This is in accordance with a non-negativity constraint for each of the terms in Eq.~\eqref{eq:delta_P1}. Note that this corollary provides a more stringent constraint than the second of Kershaw's conditions (Eq.~\ref{eq:kershaw_1}).

\subsection{The M1 approximation}

To avoid the non-negativity problem of the distribution by construction, it is necessary to formulate an evolution equation for $f_1$ that always respects the bounds of Eqs.~\eqref{eq:kershaw_1} and \eqref{eq:kershaw_2}. The M1 family of closure relations accomplishes this while simultaneously capturing the beam mode of radiation/CR transport without the addition of another evolution equation. This is accomplished by deriving theories that result in second moments that entirely depend on $f_0$ and $f_1$ while simultaneously fulfilling Kershaw's conditions in Eqs.~\eqref{eq:kershaw_0} to \eqref{eq:kershaw_2}.  In the M1 approximation, the zeroth and second moment are related by the Eddington factor $D$ via
\begin{align}
   f_2 = D f_0.
\end{align}
$D$ depends on $f_0$ and $f_1$ in such a way so that the beam transportation mode and the P1 approximation are realised in the appropriate limits.
With this definition, the evolution equations for the first two moments in Eqs.~\eqref{eq:two_moment_1} and \eqref{eq:two_moment_2} read as:
\begin{align}
    \frac{1}{c}\frac{\partial f_0}{ \partial t} &+ \bnabla \bcdot (\mathbfit{b} f_1) = \left.\frac{1}{c}\frac{\partial f_0}{\partial t} \right\vert_\mathrm{scatt}, \label{eq:M1_1}\\
    \frac{1}{c}\frac{\partial f_1}{\partial t} &+ \mathbfit{b} \bcdot \bnabla (D f_0) + (\bnabla \bcdot \mathbfit{b}) \frac{3D - 1}{2}  f_0 = \left.\frac{1}{c}\frac{\partial f_1}{\partial t} \right\vert_\mathrm{scatt}.\label{eq:M1_2}
\end{align}
The $f_1$ equation of this approximation and thus the full evolution of the two-moment system is equal to it's P1 analogue if $D = 1/3$, which is assumed in the P1 approximation. A similar result is derived by \citet{2021Hopkins}.

\subsubsection{Kershaw's closure}
The most accessible route to derive an M1-like approximation is given by the \citet{1976Kershaw} closure. In our presentation of this closure we follow the derivation of \citet{2016Schneider} who presents a theory for radiative transfer in slab geometry. Due to the similarity of the CR and radiative transfer problems, we can follow their derivation and apply it to the CR case studied here. The general idea of Kershaw's closure is to blend between a P1-type approximation and a beam approximation, which only depends on the ratio $f_1 / f_0$ and thus on the anisotropy of radiation/CR transport. In the limit $\vert f_1 / f_0\vert\ll1$, the P1 approximation should be recovered while the beam representation should be used for $\vert f_1 / f_0\vert \sim 1$.      
This behaviour is realised by the following distribution:
\begin{align}
    f_\text{Kershaw} &= \frac{1}{3} f_\text{sym} + \frac{2}{3}  f_\text{beam} , \\
    \text{where} \nonumber \\
    f_\text{sym} &= (f_0 - f_1) \, \delta (\mu + 1) + (f_0 + f_1) \, \delta (\mu - 1), \\
    f_\text{beam} &= f_0 \,\delta (\mu - f_1 / f_0).
\end{align}
The Eddington factor is readily calculated to be:
\begin{align}
D = \frac{1}{3} + \frac{2}{3} \left(\frac{f_1}{f_0}\right)^2.
\end{align}
If the Eddington factor $D=1/3$ then transport is nearly isotropic ($\vert f_1 / f_0 \vert \sim 0$), and vice versa, restoring the P1 limit. If, instead, there is an anisotropy ($\vert f_1 / f_0 \vert > 0 \rightarrow D > 1/3$) the evolution of the P1 and Kershaw's approximation for CRs will noticeably differ due the influence of the focusing term in Eq.~\eqref{eq:M1_2}. The beam transportation mode is realised for a high degree of anisotropy ($\vert f_1 / f_0 \vert \sim 1 \rightarrow D \sim 1$). In this limit, CRs are transported with characteristic speeds $\pm c$ as expected for CRs that are beamed.

\subsubsection{Levermore's closure}

Others means and arguments can be used to infer Eddington factors for radiation hydrodynamics that provide the same limits as Kershaw's closure. One possible choice of $D$ is obtained by assuming an isotropic distribution in the frame comoving with velocity $f_1 / f_0$ and employing the covariant transformation laws of energy, momentum, and pressure to get \citep{1984Levermore,2014Hanawa}: 
\begin{equation}
D = \frac{1}{3} + \frac{2 (f_1 / f_0)^2}{2 + \sqrt{4 - 3 (f_1 / f_0)^2}}. \label{eq:levermore}
\end{equation}
We see that the required limits are realised in the Levermore closure: 
\begin{align}
\vert f_1 / f_0\vert \rightarrow 0 &\text{\quad results in \quad} D  \rightarrow 1/3, \\ 
\vert f_1 / f_0\vert \rightarrow 1 &\text{\quad results in \quad} D  \rightarrow 1. 
\end{align}
Again, using the correspondence between radiative and CR transport, we can apply Levermore's closure also to the CR case. In the remainder of this paper we will exclusively use Levermore's closure in Eq.~\eqref{eq:levermore} for the M1 approximation due to it's popularity in the literature and because its derivation is based on physics arguments, i.e., we solve the CR moment equations Eqs.~\eqref{eq:M1_1} and \eqref{eq:M1_2} complemented by Eq.~\eqref{eq:levermore} for $D$.

\subsection{The H1 approximation}

A shortcoming of both, the P1 and M1 approximation, is their inability to capture two interpenetrating beams of CRs. This can be easily seen by counting the number of independent variables: each beam is defined by the number of CRs and their average velocity. For two beams this amounts to four independent quantities, which cannot be described in the P1 or M1 approximation because both models are characterised by only two independent quantities. The obvious way to cure this shortcoming is to generalise the P1 or M1 approximations by including more independent variables. Let $N$ denote an arbitrary integer that states the order of the considered polynomial, then the P$N$ approximation for the focused transport equation is given in \citet{BookZank}. Increasing the number of degrees of freedom and thus the number of moments makes the task to find a positive distribution $f$ that complies with these moments more challenging.

Instead of using the P$N$ approximation as a generalisation, here we present a new H1 approximation. Instead of increasing the order of the considered polynomial, the H1 approximation is characterised by an improved discretisation in pitch-angle space. In particular we split the $\mu$-domain in two equal parts (where the `H' in H1 denotes the half space in pitch angle) and assume that $f$ is given by a linear function in each subdomain:
\begin{align}
    f_\text{H1} = \begin{cases}
    f_0^- + 12 f_1^- (\mu + 1/2) & \text{for} \quad \mu < 0, \\
    f_0^+ + 12 f_1^+ (\mu - 1/2) & \text{for} \quad \mu > 0.  
    \end{cases}
\end{align}
Taking the $1$ and $\mu \pm 1/2$ moments of Eq.~\eqref{eq:vlasov} in the range of $\mu \in [-1,0]$ and $\mu \in [0, 1]$ yields the following evolution equations for the four quantities $f_0^\pm$ and $f_1^\pm$:
\begin{align}
    \frac{1}{c} \frac{\partial f_0^\pm}{\partial t} &+ (\mathbfit{b} \bcdot \bnabla) \left( \pm \frac{f_0^\pm}{2} + f_1^\pm \right) \nonumber \\ &+ (\bnabla \bcdot \mathbfit{b}) \left(\mp \frac{1}{2} f(0) \pm \frac{f_0^\pm}{2} + f_1^\pm \right) = \frac{1}{c} \left. \frac{\partial f_0^\pm}{\partial t} \right\vert_\text{scatt}, \\
    \frac{1}{c} \frac{\partial f_1^\pm}{\partial t} &+ (\mathbfit{b} \bcdot \bnabla) \left( \frac{f_0^\pm}{12} \pm \frac{f_1^\pm}{2} \right) \nonumber \\ 
    &+ (\bnabla \bcdot \mathbfit{b}) \left( \frac{1}{4} f(0) - \frac{f_0^\pm}{4} \pm f_1^\pm \right) = \frac{1}{c} \left. \frac{\partial f_1^\pm}{\partial t} \right\vert_\text{scatt},
\end{align}
where the value of $f(0)$ is introduced through integration by part. Formally, $f$ is not defined at $\mu = 0$. We continue by adopting the arithmetic average of the limiting values to the left and right of 0:
\begin{equation}
    f(0) = \frac{1}{2} (f_0^+ + f_0^-) - 3(f_1^+ - f_1^-).
\end{equation}
Information for this set of equations is propagated with velocities $c(-1/2\pm\sqrt{1/12})$ (associated with the fields $f_0^-$ and $f_1^-$) or $c(+1/2 \pm\sqrt{1/12})$ (associated with the fields $f_0^+$ and $f_1^+$). The fastest propagation speeds in this approximation are $(\pm 1/2 \pm \sqrt{1/12})c \approx \pm 0.78 c$. Consequently, the H1 model can be interpreted as two ultra-relativistic fluids that move with mean speeds $\pm c/2$ along the magnetic field and each experiencing a pressure with an isothermal sound speed of $\sqrt{1/12} c \approx 0.28 c$. This confirms that the H1 approximation can describe two interpenetrating streams of CRs.

By defining $f_0 = f_0^- + f_0^+$ and $\Delta f_0 = f_0^+ - f_0^-$ we can derive a conservation law for $f_0$ and a non-conservative hyperbolic differential equation for $\Delta f_0$ in the form of Eq.~\eqref{eq:vlasov2}: 
\begin{align}
    \frac{1}{c} \frac{\partial f_0}{\partial t} &+ \bnabla \bcdot \left[ \mathbfit{b}  \left( \frac{\Delta f_0}{2} + f_1^+ + f_1^- \right) \right] = \frac{1}{c} \left. \frac{\partial f_0}{\partial t} \right\vert_\text{scatt}, \label{eq:H1_1}\\
    \frac{1}{c} \frac{\partial \Delta f_0}{\partial t} &+ (\mathbfit{b} \bnabla) \bs{:} \left[ \left( \frac{f_0}{2} + f_1^+ - f_1^- \right) \mathbfss{1} \right. \nonumber \\ &\hspace{-5pt}
    + \left.\left( - f(0)   + \frac{f_0}{2} + f_1^+ - f_1^-\right) (\mathbfit{b} \mathbfit{b} - \mathbfss{1}) \right] = \frac{1}{c} \left. \frac{\partial \Delta f_0}{\partial t} \right\vert_\text{scatt}.    \label{eq:H1_2}
\end{align}
The equations for $f_1^\pm$ can be put in the same form resulting in:
\begin{align}
\frac{1}{c} \frac{\partial f_1^\pm}{\partial t} &+ (\mathbfit{b} \bnabla) \bs{:} \left[ \left( \frac{f_0^\pm}{12} \pm \frac{f_1^\pm}{2}\right) \mathbf{1} \right. \nonumber \\ &+ \left.\left(\frac{f(0)}{4} - \frac{f_0^\pm}{4} \pm f_1^\pm\right) (\mathbfit{b} \mathbfit{b} - \mathbf{1}) \right] = \frac{1}{c} \left. \frac{f_1^\pm}{\partial t} \right\vert_\text{scatt}. 
\label{eq:H1_3}
\end{align}
We use Eqs.~\eqref{eq:H1_1} to \eqref{eq:H1_3} in our simulations employing the H1 approximation. The non-negativity of $f$ in this approximation is guaranteed in all cases if 
\begin{align}
\vert f_1^- \vert < f_0^- / 6 \quad \text{ and } \quad \vert f_1^+ \vert < f_0^+ / 6.
\end{align}

\subsection{Radiation hydrodynamics}
For completeness, we now review the basic aspects of radiation hydrodynamics. The fundamental difference of CR and radiative transport is the treatment of the direction of transport. For CR transport, this direction is fixed to be the direction of the local magnetic field. By contrast, the direction of radiation transport cannot be determined a priori but is an evolving quantity. This is reflected in the way we define moments of $f$ in radiation hydrodynamics: instead of only using pitch-angle moments (scalar quantities), the three-dimensional nature of the general radiative transport equation requires us to use full directional moments (tensor quantities) with
\begin{equation}
    \mathbfit{f}_n = \frac{1}{4\uppi} \int_{S^2} \mathrm{d}^2 \Omega \, \mathbfit{d}^{(n)} f(\mathbfit{d}),
\end{equation}
where $\mathrm{d}^2 \Omega$ is the differential solid angle, $\mathbfit{d}$ is the propagation direction of radiation, $\mathbfit{d}^{(n)}$ is the $n$-th order tensor product of $\mathbfit{d}$ with itself, and $\mathbfit{f}_n$ is the $n$-th moment of $f$ and a rank-$n$ tensor. The P1 approximation for radiation hydrodynamics is readily derived by considering the distribution
\begin{equation}
    f = f_0 + 3 \mathbfit{d} \bcdot \mathbfit{f}_1,
\end{equation}
taking the $1$ and $\mathbfit{d}$ averages of the radiative transfer equation yields:\footnote{Here we omit radiation-matter interactions.}
\begin{align}
    \frac{1}{c}\frac{\partial f_0}{\partial t} &+ \bnabla \bcdot \mathbfit{f}_1 = 0 \\
    \frac{1}{c}\frac{\partial \mathbfit{f}_1}{\partial t} &+ \bnabla \bcdot (\mathbfss{D}f_0)  = 0,
\end{align}
where the P1 Eddington tensor is given by:
\begin{equation}
    \mathbfss{D} = \frac{1}{3}\, \mathbf{1}.
\end{equation}
The M1 approximation for radiation hydrodynamics replaces this tensor with \citep{1984Levermore}:
\begin{equation}
    \mathbfss{D} = \frac{3D - 1}{2} \frac{\mathbfit{f}_1}{f_1} \frac{\mathbfit{f}_1}{f_1} + \frac{1-D}{2} \mathbfss{1}.
\end{equation}
where $D$ is given by Eq.~\eqref{eq:levermore}.

\subsection{CR scattering}
\label{sec:scattering}

So far we only accounted for the interaction of CRs with a large scale mean magnetic field but there are several mechanisms, which produce magnetic field fluctuations with typical length scales comparable to the gyroradius of CRs. Here, we focus on the interaction of CRs with small-scale gyroresonant Alfv\'en waves mediated by pitch-angle scattering. Other mechanisms such as external confinement of CRs by MHD turbulence \citep{2006Lazarian} or scattering of CRs by the intermediate scale instability \citep{2021Shalaby} are not considered but could be included in an extension of the presented theory. The phase-space diffusion coefficients for pitch-angle scattering in our mono-energetic approximation can be derived from \citet{1989Schlickeiser} in the first-order limit in $\varv_\mathrm{a} / c$ so that the scattering term reads:
\begin{align}
    \left. \frac{\partial f}{\partial t} \right\vert_\text{scatt} = \frac{\partial}{\partial \mu} \left\lbrace \frac{1 - \mu^2}{2}  \nu_\pm \left[ \left(1 \mp 2 \mu \frac{\varv_\mathrm{a}}{c} \right) \frac{\partial f}{\partial \mu} \mp 3 \frac{\varv_\mathrm{a}}{c} f \right] \right\rbrace, \hspace{-8pt} \label{eq:scattering}
\end{align}
where $\nu_\pm = \nu_\pm(\mu)$ is the scattering rate of CRs with gyroresonant Alfv\'en waves that travel in the direction of ($+$) or against ($-$) the magnetic field with Alfv\'en velocity $\bs{\varv}_\mathrm{a}$. This first-order limit in $\varv_\mathrm{a} / c$ is sufficient because accounting for second-order terms is only necessary to ensure energy conservation \citep{2019Thomas}. Energy is naturally conserved in our mono-energetic approximation and thus this low-order approximation can be used to simplify the calculations. The actual value of the scattering coefficient is set by the energy balance of magnetic fluctuations of gyroresonant Alfv\'en waves. In general, this scattering term is actually a sum over the scattering contributions from both types of Alfv\'en waves. We suppress this sum for readability and in our, yet to be presented, simplified model where at most one type of Alfv\'en waves provides the scattering.

One of the main sources of energy is the CR streaming instability \citep{1969Kulsrud} that is active once CRs stream faster than Alfv\'en waves (approximately when $\vert \partial_\mu f \vert > (\varv_\mathrm{a} / c) f$). This converts kinetic energy from CRs to energy contained in Alfv\'en waves which in turn provide the scattering agents of CRs. Damping of Alfv\'en waves is provided by various mechanisms \citep{2013Zweibel} and thermalises this energy. Typically, the growth rate of Alfv\'en wave energy and the total damping rate balance each other to reach a quasi-steady state.

We are not able to follow this energy balance in our idealised model for CR transport here. This is mostly due to our assumption that CRs are mono-energetic and thus cannot lose energy. Nevertheless, to mimic the behaviour of CR losses, we assume that the scattering frequency is non-vanishing if we were to model the energy transfer in a more complete description and energy would be transferred from the CRs to the gyroresonant Alfv\'en waves. \citet{2019Thomas} observed that for the gyroresonant interaction, we have
\begin{equation}
    \text{CR energy loss} \sim \varv_\mathrm{a} \times\text{ CR momentum loss,}
\end{equation}
which is positive once CRs travel faster (defined in by some average) than Alfv\'en waves and are decelerated towards the Alfv\'en velocity. Because CR energy loss equals the gain of Alfv\'en wave energy, we assume that gyroresonant Alfv\'en waves are present and scatter CRs once CRs have been decelerated by Alfv\'en waves.

\subsubsection{CR scattering in the P1 and M1 approximation}

A discussion of gyroresonant pitch-angle scattering in a fully energy-dependent setting using the P1 approximation is presented by \citet{2019Thomas}. Here, we use a more simplified model for the P1 and M1 approximation. We assume that the scattering rates are independent of pitch angle:
\begin{equation}
    \nu_\pm(\mu) = \nu_\pm = \text{const}. 
\end{equation}
Taking the $1$ and $\mu$ moments in the P1 approximation yields the following expressions for the scattering terms in Eqs. ~\eqref{eq:P1_1} and \eqref{eq:P1_2}:
\begin{align}
 \left.\frac{1}{c}\frac{\partial f_0}{\partial t} \right\vert_\mathrm{scatt} &= 0, \\
 \left.\frac{1}{c}\frac{\partial f_1}{\partial t} \right\vert_\mathrm{scatt} &=- \frac{\nu_\pm}{c} \left(f_1 \mp \frac{\varv_\mathrm{a}}{c} f_0 \right),
\end{align}
while the result for the M1 approximation in Eqs. ~\eqref{eq:M1_1} and \eqref{eq:M1_2} is given by:
\begin{align}
 \left.\frac{1}{c}\frac{\partial f_0}{\partial t} \right\vert_\mathrm{scatt} &= 0, \\
 \left.\frac{1}{c}\frac{\partial f_1}{\partial t} \right\vert_\mathrm{scatt} &= -\frac{\nu_\pm}{c} \left(f_1 \mp \frac{\varv_\mathrm{a}}{c} \frac{1 + 3 D}{2} f_0 \right).
 \label{eq:M1_scatt}
\end{align}
Because $D = 1/3 + \mathcal{O}\left((f_1/f_0)^2\right)$, the scattering terms of the P1 and M1 approximation only differ if the flow of CRs is highly anisotropic. Thus, Eq.~\eqref{eq:M1_scatt} reduces to its P1 counterpart if $\vert f_1 / f_0 \vert \to 0$, as expected. For comparability and simplicity, we assume that the same expressions for the activation of the scattering holds true for the P1 and M1 approximations. For both approximations, CRs are decelerated once $\vert f_1 \vert > (\varv_\mathrm{a} / c) f_0$ and thus we use
\begin{equation}
    \nu_\pm \begin{cases}
    > 0 & \quad\text{ if } f_1 \gtrless \pm (\varv_\mathrm{a} / c) f_0,
 \\
 = 0 &\quad\text{ else.} \end{cases}
\end{equation}
This procedure only describes the activation of scattering. The actual value used for the scattering frequency (once it is activated) is assumed to be a constant and given for our numerical experiment in Section~\ref{sec:numerics}.

\subsubsection{CR scattering in the H1 approximation}

For the H1 approximation, we include a dependence of scattering coefficients on $\mu$ and assume that they have the following form:
\begin{align}
\nu_\pm = \begin{cases}
 \nu^\rmn{L}_\pm & \quad \text{for} \quad \mu < 0, \\
 \nu^0_\pm & \quad \text{for} \quad \mu = 0, \\
 \nu^\rmn{R}_\pm & \quad \text{for} \quad \mu > 0. \\
\end{cases}
\end{align}
The reasoning for this splitting is apparent once the four scattering source terms are calculated: to derive these terms for $f_0$ and $\Delta f_0$, we calculate the 1-moments on the intervals $[-1,0]$ and $[0,1]$ to get:
\begin{align}
 \left.\frac{1}{c}\frac{\partial f_0}{\partial t} \right\vert_\mathrm{scatt} &= 0, \\
 \left.\frac{1}{c}\frac{\partial  \Delta f_0}{\partial t} \right\vert_\mathrm{scatt} &= -\frac{\nu_\pm(0)}{c} \left(\frac{\partial f}{\partial \mu} (0) \mp 3 \frac{\varv_\mathrm{a}}{c} f(0) \right).
\end{align}
The scattering rate $\nu_\pm^0(0)$ thus mediates the scattering of CRs through the $\mu=0$ point. Without this term, CRs would not be able to scatter through this point, which is a problem commonly referred to as the $90^\circ$ problem. Numerical studies \citep{2019Bai} and more detailed theoretical models for pitch-angle scattering \citep{2005Shalchi} indicate that while scattering at $\mu=0$ is reduced, it is sufficiently frequent and scatters CRs between the two hemispheres. Moving on, we calculate the $\mu$ moments on the intervals $[-1,0]$ and $[0,1]$ to get:
\begin{align}
 \left.\frac{1}{c}\frac{\partial f_1^-}{\partial t} \right\vert_\mathrm{scatt} &= \frac{\nu_\pm(0)}{4c}  \left(\frac{\partial f}{\partial \mu} (0) \mp 3 \frac{\varv_\mathrm{a}}{c} f(0) \right) \nonumber \\
 &\phantom{=} -\frac{\nu_\pm^\rmn{L}}{c} \left[ 4f_1^- \mp \frac{\varv_\mathrm{a}}{c} \left(f_0^- -\frac{3}{2}f_1^-\right)\right], \\
 \left.\frac{1}{c}\frac{\partial f_1^+}{\partial t} \right\vert_\mathrm{scatt} &= \frac{\nu_\pm(0)}{4c}  \left(\frac{\partial f}{\partial \mu} (0) \mp 3 \frac{\varv_\mathrm{a}}{c} f(0) \right) \nonumber \\
 &\phantom{=} -\frac{\nu_\pm^\rmn{R}}{c} \left[ 4f_1^+ \mp \frac{\varv_\mathrm{a}}{c} \left(f_0^+ +\frac{3}{2}f_1^+\right)\right].
\end{align}
Thus, $\nu_\pm^\rmn{L}$ describes the scattering of $f_1^-$ while $\nu_\pm^\rmn{R}$ describes the scattering of $f_1^+$.
These expressions contain $f$ and $\partial_\mu f$ evaluated at $\mu = 0$, which is formally not well defined. We follow the approach of \citet{2005vanLeer}, who consider discontinuous Galerkin discretisations of diffusion equations, and derive:
\begin{align}
    f(0) &= \frac{1}{2} (f_0^+ + f_0^-) - \frac{1}{2} (f_1^+ - f_1^-), \\
    \frac{\partial f}{\partial \mu}(0) &= \frac{9}{4}(f_0^+ - f_0^-) - \frac{15}{2}(f_1^+ + f_1^-)
\end{align}
by fitting a forth-order polynomial through the four independent variables. While this approach has been derived for a numerical method, it is appropriate here because the presented H1-approximation can be interpreted as a discontinuous Galerkin approximation on two computational cells with linear basis functions. Again, we consider scattering, described by one of the three rates, to be active once their scattering induces a momentum loss of CRs:
\begin{align}
    \nu_\pm^\rmn{L} & \begin{cases}
    > 0 & \quad\text{ if } 4f_1^- \gtrless \pm \frac{\dps\varv_\mathrm{a}}{\dps c} \left(f_0^- - \frac{\dps 3}{\dps 2} f_1^-\right),
 \\
 = 0 &\quad\text{ else,} \end{cases} \\
    \nu_\pm^0 & \begin{cases}
    > 0 & \quad\text{ if } \partial_\mu f(0) \gtrless \pm 3 \frac{\dps \varv_\mathrm{a}}{\dps c} f(0),
 \\
 = 0 &\quad\text{ else,} \end{cases} \\
    \nu_\pm^\rmn{R} & \begin{cases}
    > 0 & \quad\text{ if } 4f_1^+ \gtrless \pm \frac{\dps\varv_\mathrm{a}}{\dps c} \left(f_0^+ + \frac{\dps 3}{\dps 2} f_1^+\right),
 \\
 = 0 &\quad\text{ else.} \end{cases}
\end{align}
This concludes our derivation of the theoretical models. 

\section{Numerical Example}
\label{sec:numerics}

We continue by describing our numerical discretisation and simulation setup. We present a suite of simulations that all use the same initial conditions and parameters but employ different transport methods. We start by considering radiation transport and compare the P1 and M1 fluid approximations to a more accurate discrete ordinates solution. By allowing transport solely along magnetic fields in the same numerical setup, we compare the P1, M1 and H1 fluid models for CR transport. In particular, we compare simulations with and without CR scattering to study the differences between the fluid approximations.

\subsection{Numerical method and setup}

In our simulations we use a two-dimensional Cartesian grid in the computational domain $x, y \in [0, 1]^2$ with $1024^2$ equally spaced resolution elements. We employ a standard finite volume method to evolve the initial conditions until $t=2$ or two light crossing times (with $c=1$). The time integration is accomplished using Heun's second-accurate Runge-Kutta method where the length of each timestep is given by $\Delta t = 0.2 \Delta x / c$ and $\Delta x$ denotes the side length of the resolution elements. The finite volume scheme uses a piecewise linear reconstruction of the primitive variables $f_0$, $f_1 / f_0$, $\mathbfit{f}_1 / f_0$, $f_0^\pm$ and $f_1^\pm / f_0^\pm$ (where applicable) onto the center of interfaces between resolution elements using the monotonised central slope limiter \citep{1979vanLeer}. At the interface we use approximate Riemann solvers to calculate the fluxes exchanged between neighbouring resolution elements. For simulation of radiation hydrodynamics using the P1 closure we use the Lax-Friedrichs Riemann solver; for simulations with the M1 closure we use the HLLC Riemann solver of \citet{2007Berthon}. We also include a simulation with a discrete ordinates approximation of the radiative transfer equation \eqref{eq:boltzmann} where we simulate 200 discrete rays, equally spaced in all directions of the plane, per computational cell and call this solution S200. Radiation in each ray is upwinded at interfaces. For simulations of CR hydrodynamics using the P1 closure we use the localised Lax-Friedrichs Riemann solver presented in \citet{2021Thomas}; for simulations with the M1 or H1 closure we use a straightforward extension of the aforementioned Riemann solver that accounts for the varying Eddington factor in the M1 case or for the four independent quantities in the H1 case. All boundaries of the computational domain are periodic.

Radiation or CRs are set up using a random population of cloudlets. We place 30 circular cloudlets with random radii $r$ uniformly distributed in $[0, 0.1]$. The center of the cloudlets are distributed uniformly throughout the computational domain. We add  for each cloudlet a cloud contribution of
\begin{align}
    f_{0, \text{cloudlet}} = \begin{cases}
    0.1 & \text{if }  \vert\vert \mathbfit{x} - \mathbfit{x}_\mathrm{c} \vert\vert < r, \\
    0 & \text{otherwise,}
    \end{cases}
\end{align}
where $\mathbfit{x}_\mathrm{c}$ is the center of the cloudlet and adopt a background of radiation or CRs given by  $f_{0,\text{background}} = 10^{-3}$. Every other moment, e.g. $f_1$, $\mathbfit{f}_1$, $f_1^\pm$  and $\Delta f_0$ are initially set to zero if the transport model follows their evolution. For the S200 simulation, radiation is distributed equally among all rays.

We use random instead of handpicked initial conditions to demonstrate the different behaviours of the respective transport models and to highlight their differences while reducing possible biases introduced by the initial conditions. These are chosen to trigger the interaction between radiation or CRs emerging from different cloudlets so that the resulting evolution is non-trivial.

Our CR hydrodynamics simulations contain a magnetic field that we initialize with a magnetic vector potential $\mathbfit{A}$ so that we obtain a divergence-free magnetic field via $\mathbfit{B} = \bnabla \btimes \mathbfit{A}$ using second-order finite differences. We set the in-plane components of the vector potential to zero: $A_x = 0$ and $A_y = 0$. The out-of-plane component $A_z$ is initialized in a similar fashion as $f_0$: we place 100 magnetic loops randomly in the whole domain whereby each loop is set up via:
\begin{align}
    A_{z, \mathrm{loop}} = \begin{cases} 
    0.2 - \vert\vert \mathbfit{x} - \mathbfit{x}_\mathrm{c} \vert\vert & \text{if } \vert\vert \mathbfit{x} - \mathbfit{x}_\mathrm{c} \vert\vert < 0.2,  \\ 0 & \text{otherwise,} \end{cases} 
\end{align}
where $\mathbfit{x}_\mathrm{c}$ is the center of the magnetic field loop.
The resulting vector potential is given by the sum of the individual contributions $A_{z, \mathrm{loop}} $ of all magnetic loops. The high number of magnetic loops is required to guarantee a complete covering of the computational domain with magnetic loops and to get a resulting irregular topology in the magnetic field. 

The remaining free parameters in our setup are $c$, $\varv_\mathrm{a} / c$ and the scattering frequencies. We chose $c=1$ for all simulations and $\varv_\mathrm{a} / c = 0.1$ for simulations with CR scattering. This is a conservative high value as in the interstellar, the circumgalactic, and the intracluster medium typical values have $\varv_\mathrm{a} < 1000~\mathrm{km} ~ \mathrm{s}^{-1}$. We use this high value to place an upper bound on the differences between the different transport methods. In addition, most numerical codes with two-moment CR hydrodynamics capabilities also employ a reduced speed of light approximation where $c$ is artificially lowered to reduce numerical diffusion and computational cost \citep{2018Jiang, 2021Thomas}. If CR scattering is activated, we adopt $\nu = 1000$ with $\nu$ representing $\nu_\pm$, $\nu^\rmn{L}_\pm$, $\nu^0_\pm$ or $\nu^\rmn{R}_\pm$. This value is based on the following consideration: for interstellar and circumgalactic medium applications the light crossing time across 1~kpc is $t_\mathrm{cross} \sim \mathrm{kpc} / c = 10^{11}~\mathrm{s}$ while the typical timescale of CR scattering is $t_\mathrm{scatt} \sim 3 \kappa / c^2 = 10^{8}~\mathrm{s}$ where we assume that the diffusion coefficient of GeV CRs is $\kappa = 3\times10^{28}~\rmn{cm}^2~\rmn{s}^{-1}$. Fixing the ratio $t_\mathrm{cross} / t_\mathrm{scatt}$ yields for $t_\mathrm{cross} = 1$ in our simulations $t_\mathrm{scatt} = 1 / \nu = 1/1000$.

We explicitly enforce the positivity bounds $\vert f_1 \vert < f_0 / \sqrt{3}$ for the P1 model and $\vert f_1^\pm \vert < f_0^\pm / 6$ for the H1 model by clipping $f_1$ or $f_1^\pm$ to the maximum allowed value if they violate this limit at the beginning of a timestep in the simulation. The same is done for $\mathbfit{f}_1$ in the radiation P1 model where we change only the magnitude of $f_1$ but not its direction.
We checked for numerical convergence of the presented simulations by running each simulations at half the resolution and found no significant differences between the results at both resolutions. To check for numerical convergence of the discrete ordinate S200 simulation with respect to the number of radiation rays used, we increased the number of rays from 200 to 400 and found no significant differences of the results.

\begin{figure*}
	\includegraphics[width=\linewidth]{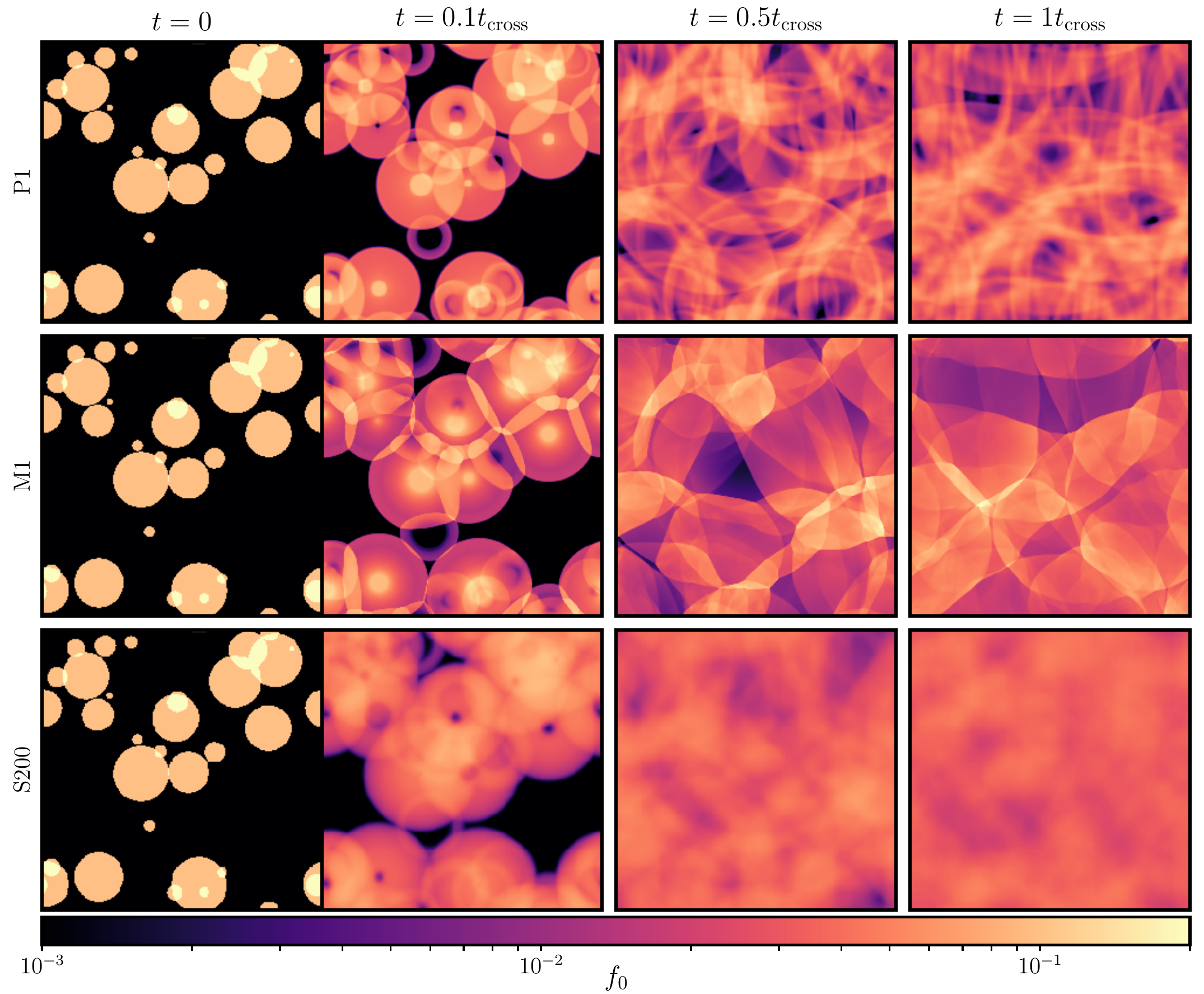}
	\caption{The first moment $f_0$ (or photon number density) at different times as calculated with the P1 and M1 fluid models and the S200 discrete ordinates solution. We display the entire simulation box. }
	\label{fig:radiation}
\end{figure*}

\subsection{Radiation hydrodynamics}
\label{sec:radiation}

In our first set of simulations, we compare the evolution of radiation cloudlets. As we do not account for any radiation-matter interaction in our simulations, transport of radiation is always in the free-streaming regime and radiation should be transport at its maximum velocity $c$ freely without any interactions at all. Formally, no hydrodynamic description for radiation should be applied to these simulations at all because this approximation is expected to break down. Nevertheless, the results of these simulations are instructive as they enable us to gauge the maximum influence of closure relations in a worst-case scenario. This demonstrates the wide-ranging implications that the choice of a transport model can have.  

The results are presented in Fig.~\ref{fig:radiation}. At early times ($t=0.1t_\mathrm{cross}$), the differences between the P1 and M1 approximations appear to be minor. The cloudlets expand mostly radially. The major difference between the two approximations is how the radial expansion shapes a cloudlet. In the P1 approximation, the expansion of each cloudlet is uniform inside an annulus that starts at the edge of the cloudlet and extents into is interior via an expansion front. The radiation inside this annulus is expanding uniformly as can be seen in the homogeneous brightness of such annuli. For the M1 approximation, the propagation model is different. Here, the cloudlets are not uniformly expanding but are rarefied and develop smooth gradients from the central brightness towards the edge of these expanding cloudlets. For both transport models, the expansion of a cloudlet changes its morphology into a ring-like structure for isolated cloudlets. Large cloudlets have a brightness maximum at their geometric centre. This is not caused by a built-up of radiation but is due to a lag in expansion. The centre of those cloudlets was not hit by the expansion wave and is, consequently, not expanding.

For cloudlets with nearby neighbours an additional difference can be observed between the P1 and M1 closure relations. At points where the expansion fronts of two cloudlet overlap, the radiation density in the P1 model results in a superposition of photon packages while in the M1 model  shock fronts emerge and change the propagation direction of the radiation flux into non-radial directions. Because of the rarefactions of radiation and the developing shock fronts, the M1 closure model exhibits characteristics of \textit{fluidisation} of radiation. This self-interaction of radiation is not induced by any physical process but a manifestation of the fluid approximation and must be regarded as a short-coming of this method.

Both, the P1 and the M1 approximations compare poorly to the discrete ordinates solution. In this model, expanding cloudlets show smoother outer edges and no central brightness enhancement. The smoothness of the evolution can be attributed to the local non-radial propagation of radiation that is possible in the S200 model. Nevertheless, the global radial expansion of a cloudlet is caused by the overlap of radiation propagating into different, non-radial directions. Furthermore, we do not observe central brightness maxima in the S200 model because this model does not feature any expansion fronts that first need to travel inside a cloudlet in order to trigger an expanding motion. Instead, the entire cloudlets starts to expand with the beginning of the simulation.

The expansion velocity of the cloudlets is different for the various transport models. For every model this velocity is equal to the fastest information propagation velocity. Hence, the expansion velocity is $c / \sqrt{3}\approx0.57c$ for the P1 approximation while it is $c$ for the M1 and S200 approximations. This leads to smaller expanding cloudlets in the P1 case in comparison to the expanding cloudlets in the M1 and S200 cases, which share the same spatial extent.

At later times, at $t=0.5 t_\mathrm{cross}$ and $t=1 t_\mathrm{cross}$, we do not find similarities between the presented transport methods. The mechanisms driving the differences at earlier times are more pronounced because the radiation that originates from multiple individual cloudlets interacts so that there is no cloudlet evolving in isolation any more. The simulations with the P1 closure shows a complex network of overlapping radiation rings resembling caustics. By contrast, streams of radiation frequently collide in the M1 simulations developing a pattern that is reminiscent of super-sonic turbulence with multiple and connected shock fronts travelling through the simulation domain. Radiation in the S200 model is diluted to the point that the computational domain is filled by a diffusive fog of radiation with no distinctive morphology.  

\subsection{CR hydrodynamics}

\begin{figure*}
	\includegraphics[width=\linewidth]{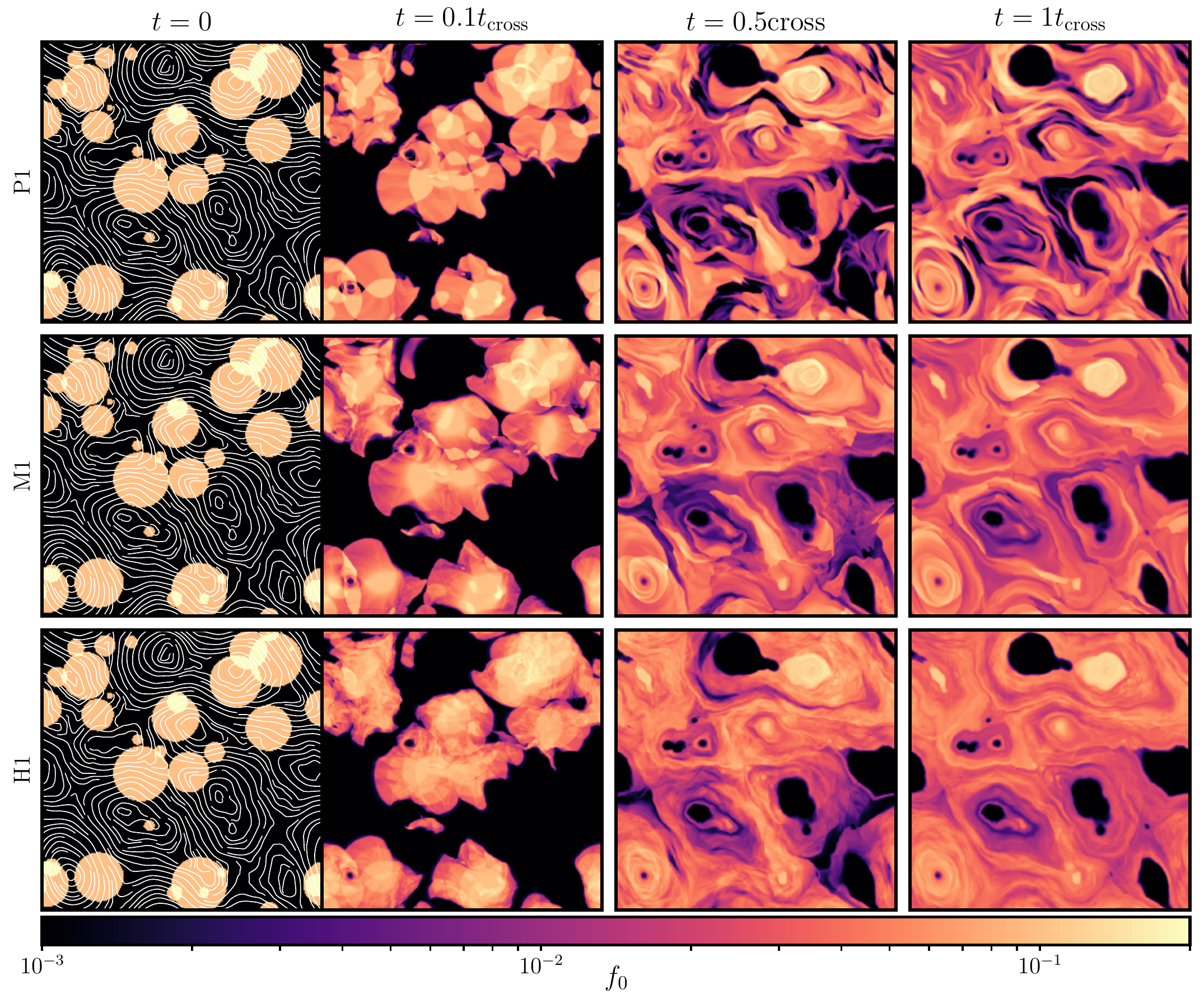}
	\caption{The first moment $f_0$ (or CR number density) at different times as calculated with the P1, M1 and H1 fluid models {\em without} CR scattering. We display the entire simulation box. The white lines in the left-most panels trace the magnetic field.}
	\label{fig:mag}
\end{figure*}

For our next set of simulations, we restrict the transport of radiation solely along magnetic fields and thus switch our attention to CR transport. In this case, no discrete ordinates solution is available as a reference solution and we need to compare the three fluid approximations (P1, M1, H1) to each other. The reason for the development of our new H1 approximation in Section~\ref{sec:hydrodynamics} is that it provides a more accurate representation of CR transport and enables us to evaluate the performance of the P1 and M1 methods. Like the previous set of simulations, here we neither account for interactions of CRs with the surrounding plasma nor with electromagnetic waves but only allow for their anisotropic transport along magnetic fields. Consequently, CRs are in the free-streaming limit at all times. We present the results of these simulations in Fig.~\ref{fig:mag}.

The structure of the P1 solution is dominated by cloudlets that are stretched and distorted while CRs are transported along the magnetic field. This leads to a patchy morphology of $f_0$ visible at $t=0.1t_\mathrm{cross}$ and $t=0.5t_\mathrm{cross}$. At later times at $t=1t_\mathrm{cross}$, the cloudlets are stretched to wrap around magnetic loops and overlap with themselves or with neighbouring cloudlets. We can understand the reason for this evolution, by realising that each cloudlet in the radiation simulation in Section~\ref{sec:radiation} developed into an expanding annulus. In the present simulation, these annuli can only expand and move along magnetic field lines. Therefore, every cloudlet will split into two parts: one moving along and the other one moving opposite to the direction of the magnetic field permeating the cloudlet. In each of these two parts, $f_0$ is rather uniform so that overlapping parts from different cloudlets create the observed patchy structure where $f_0$ increases stepwise towards local brightness maxima. This is different for the M1 model. Radiation cloudlets in the M1 model are rarefying. This is also the case for CRs transported with the M1 model: there are fewer patches with $f_0 \sim \mathrm{const.}$ present in the M1 solution in comparison to its P1 counterpart but instead more regions in which $f_0$ shows smooth gradients. The same holds true for the simulation with the H1 approximation. In the M1 and H1 models, the simulations contain regions with multiple wavefronts that cross through each other. Most importantly, in all methods, CRs are confined to magnetic islands, which we identify with closed magnetic loops with a small spatial extent, and which fill their hosting island mostly homogeneously. Regions which are not connected to a cloudlet through a magnetic field line do not contain CRs that any time. This creates the dark magnetically insulated islands in Fig.~\ref{fig:mag} whose surroundings get populated by CRs over time while no CR enters the magnetic island itself.

The differences between the results of the P1 and M1 approximations are smaller for anisotropic CR transport in comparison to radiation transport (cf.\ Fig.~\ref{fig:radiation} vs.\ Fig.~\ref{fig:mag}): in the case of radiation transport, both approximations lead to fundamentally different solutions while we can identify the most prominent features in both approximations if the transport is restricted to be along magnetic field lines. This is because CRs trace the magnetic field topology and thus features contained in the magnetic field will be inevitably present also in the global CR distribution. Only if we compare the P1 and M1 approximations along a single magnetic field line, we are able to quantify differences.

CR cloudlets expand with the same velocity as their radiation counterparts. Like in the radiation case, the M1 cloudlets expand initially with velocity $c$ along magnetic field lines while CRs in the P1 case travel at maximum speeds of $c / \sqrt{3}\approx0.57c$. The expansion velocity of CR cloudlets in the H1 approximation is placed in between the P1 and M1 case with a maximum velocity of $(1/2 + \sqrt{1/12}) c\approx0.78c$. The spherical bright spots observed at the centres of radiation cloudlets in the P1 and M1 approximations are still present for the CR case. Here, cloudlets do not radially expand with respect to a cloudlet centre but propagate along the magnetic field lines permeating the cloudlet. Because these magnetic field lines are non-uniform, the bright spots of the cloudlets develop irregular shapes.

Morphologically, we find more structure in the P1 results, which are smoothed out in case of the M1 or H1 solutions, the latter of which shows the most diffusive appearance. This is due to the increasing dynamical flexibility of the M1 approximation or the two additional degrees of freedom in the H1 approximation in comparison to P1. Overall, the M1 and H1 results resemble each other while the P1 morphology shows the largest contrast to that of the two other methods \citep[see also][for a P1-M1 comparison]{2021Hopkins}. 

In the case of the M1 and H1 methods, the magnetic focusing term provides a source of indirect scattering that can explain the similarity of both approximations. The reason why magnetic focusing can act as a scattering agent is particular easy to understand for the M1 model; in this case, the magnetic focusing term in the $f_1$-equation is given by
\begin{equation}
    \left.\frac{\partial f_1}{\partial t}\right\vert_\mathrm{focusing} = (\bnabla \bcdot \mathbfit{b}) \frac{3D - 1}{2}
\end{equation}
and $3D - 1 > 0$ for $\vert f_1 / f_0 \vert > 0$. For a given magnetic field line, CRs will be transported through regions where $\bnabla \bcdot \mathbfit{b} > 0$ and further down this field line through regions where $\bnabla \bcdot \mathbfit{b} < 0$ provided the magnetic field topology is sufficiently irregular. Depending on this sign the focusing term switches its behaviour, i.e. the CR flux $f_1$  increases or decreases. If the sign of $\bnabla \bcdot \mathbfit{b}$ is permanently changing, the evolution of $f_1$ will become more random. On the other hand, if $\bnabla \bcdot \mathbfit{b}$ does not change its sign, $f_1$ might evolve towards a state where Kershaw's condition in Eq.~\eqref{eq:kershaw_1} is violated. Whether it is possible to construct such magnetic field topologies will be left as an open question for future studies. 

Clearly this scattering effect is only important if it is the most dominant source of scattering which is the case in the presented idealised simulations in which we explicitly do not account for any other source of scattering. We also note that the resulting effect of this scattering might be further exaggerated by our fluid models. The magnetic focusing terms originates in the corresponding term in Eq.~\eqref{eq:vlasov} that is clearly not describing a scattering process but advection of $f$ through pitch-angle space. By its nature, advection is localised in pitch-angle space and transfers information contained in $f$ from $\mu$ to some neighbouring $\mu + \mathrm{d}\mu$. Because pitch-angle moments involve integrals over $\mu$ space, low-order fluid models lost most of their memory about localised information of $f$ and only retain information of the global moment structure of $f$. Consequently, fluid models cannot accurately describe the magnetic focusing process in all details. Yet, because we account for the focusing term in deriving the evolution equations for the moments $f_n$, we simultaneously account for its global effect on $f$ which thus retains some indirect memory through the scattering process as explained above.

\begin{figure*}
	\includegraphics[width=\linewidth]{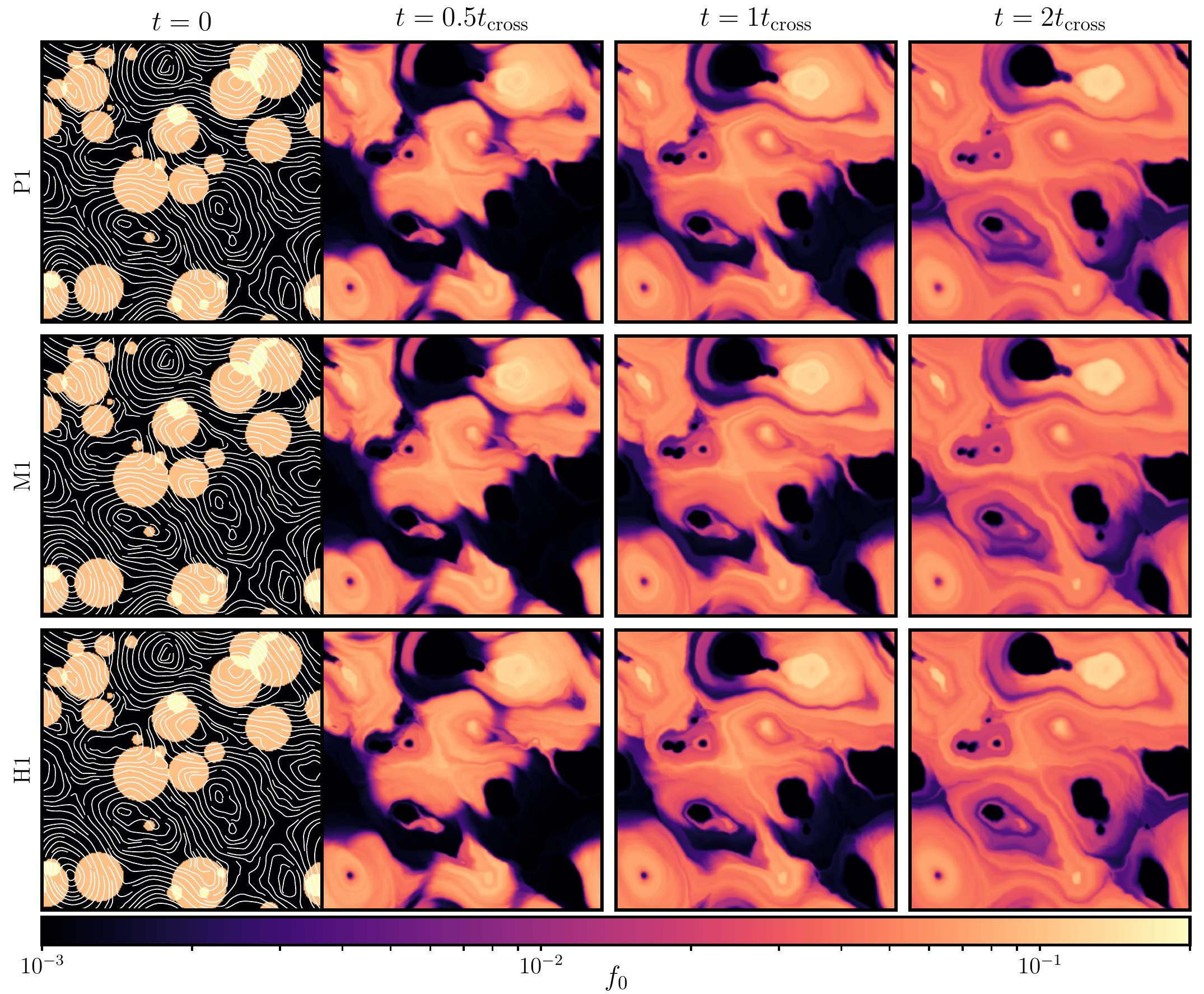}
	\caption{The first moment $f_0$ (or CR number density) at different times as calculated with the P1, M1 and H1 fluid models {\em with} CR scattering. We display the entire simulation box. The white lines in the left most panels trace the magnetic field.}
	\label{fig:crhd}
\end{figure*}

\subsection{CR hydrodynamics with scattering}

For our last set of simulations, we include CR scattering through the models described in Section~\ref{sec:scattering}. They are accurate to first order in $\varv_\rmn{a}/c$ and as such, they are comparable to each other and consistently formulated so that our results are not influenced by any differences between them. This allows us to quantify the impact of the different transport models in the case of effective CR scattering. The results of the simulations are presented in Fig.~\ref{fig:crhd}.

By including CR scattering in our simulations, the local flux of CRs is reduced, which results in a slower transport of CRs through the simulation domain. For this reason we show $f_0$ in Fig.~\ref{fig:crhd} at later times in comparison to the previous Figs.~\ref{fig:radiation} and \ref{fig:mag}. The CR dynamics slightly differs in comparison to the case without CR scattering: cloudlets smoothly spread along magnetic field lines. Sharp transitions as observed in the simulations without CR scattering are not visible. Structures with a larger contrast are caused by neighbouring magnetic fields where one field line is connected to a cloudlet and can be populated by CRs while the other is not connected and remains devoid of CRs. Neither patchiness nor rarefied structures, which are a characteristics for the P1 and M1 method of CR transport without scattering are present in this case. At late times, cloudlets dissolve into each other and build up a rather homogeneous sea of CRs. There are smooth transitions between regions filled with abundant CRs and those that contain fewer CRs (or which have not yet been reached by any) if they are connected through a magnetic field line.

There are no visible differences between the results calculated with the P1 and M1 approximations at any time. The reason for this is that the difference between both models is the way they treat CR fluxes that are mildly to fully anisotropic, i.e., $f_1 \sim f_0$. In such situations, the M1 Eddington factor $D \neq 1/3$ and differs from its P1 counterpart. However, in the presented scenario, CR scattering is effective and leads to low anisotropies in the CR distribution with $f_1 \sim (\varv_\mathrm{a} / c) f_0 = 0.1 f_0$ and thus $D \sim 1/3$ even in the M1 case. The M1 approximation is effectively reduced (by removing the main difference) to the P1 approximation if scattering removes anisotropy or, in other words, is reducing the flux of CRs to non-relativistic values. 

All major features observed in the two-moment approximations (P1 and M1) can be readily found in the simulations with the H1 approximation. However, there are minor differences between the P1/M1 and H1 results. In particular, the H1 simulation exhibits small details that are absent in the P1 and M1 cases. These are best observed in regions into which CR have just entered, i.e., where $f_0$ is small. The reason for this is that the H1 approximation provides four degrees of freedom and thus allows for more hydrodynamical waves to travel inside the simulation which then create the observed fine-grained structure. 

\section{Discussion and Conclusion}
\label{sec:discussion}
In order to investigate the differences between closure relations for CR hydrodynamics, we (i) reviewed the derivation of the P1 and M1 fluid approximations starting from the general focused transport equation, (ii) discussed how a simplified model for CR scattering with Alfv\'en waves can be implemented and (iii) presented a set of numerical simulations that highlight the deviations of the different approximations. In order to check the correctness of these approximations, we develop and present the new H1 fluid approximation which evolves four moments of the CR pitch-angle distribution instead of the two moments used by the P1 and M1 approximations. Because the P1 and M1 models were originally developed in the context of radiation hydrodynamics we also include simulations and derivations of both models in this limit and compare simulation results to a more correct radiation transfer simulation using the discrete ordinates approximation. Because (GeV or higher-energy) CRs and radiation are both relativistic (particle) species with a negligible rest mass energy density, any insight gained from the radiation case can be used to understand the CR transport.

As we advance to ever-increasing details and include more CR physics, we observe that the differences between the presented hydrodynamical models get smaller. While the P1 and M1 approximations generate large differences in the evolution of radiation, including a large-scale magnetic field for CR transport already results in a more similar spatial CR distribution in the P1 and M1 approximations because CRs are necessarily confined to and transported along magnetic field lines. The large scale structure of the resulting CR distribution is now set by the magnetic field topology, which effectively reduces the degrees of freedom for the transport and thus the dimensionality of the solution space. Nevertheless, if we compare the CR evolution along an individual magnetic field line in both approximations, we still observe differences. 

This situation changes once we include the effects of CR scattering. In this more realistic case the differences between the approximations are significantly reduced to the point where they are merely of academic interest. This is an expected result because scattering reduces the local flux so that the bulk of CRs is transported with non-relativistic velocities and the relativistic corrections of the M1 method become ineffectual. As a result, simulations with the M1 approximation are effectively indistinguishable from their P1 counterpart. Even if we consider the higher-order accurate H1 approximation, the differences between the results of the P1, M1 and H1 models are very small and have negligible astrophysical consequences. This is because CR scattering is a pitch-angle diffusion process that reduces the amount of information encoded in the pitch-angle distribution of CRs. Higher-order methods such as the H1 approximation strive to encode more information in the CR pitch-angle distribution. However, because there is very little additional fine-structure contained in the pitch-angle distribution (at timescales comparable or longer than a light crossing time) modelling these degrees of freedom does not significantly improve the accuracy of the solution.

The advantage of the M1 approximation in comparison to the P1 approximation is that it allows for the beam transport mode in a two-moment fluid description. The beam mode is characterised by a high degree of anisotropy in the CR distribution. However, if this anisotropy is large enough, additional micro-scale plasma processes such as the non-resonant hybrid instability \citep{2004Bell} and the intermediate-scale instability \citep{2021Shalaby} start to dominate and influence the CR evolution. Therefore, to provide a consistent description of CR dynamics these additional kinetic processes need to be incorporated in the fluid description through the scattering terms. Note that both instabilities act in such a way that they drive the CR distribution towards isotropy and thus towards the applicability of the regime of the P1 approximation. Lastly, we want to stress that this discussion and hydrodynamical models in general are only valid on scales significantly larger than the CR gyroradius and mean free path. On scales comparable to the CR gyroradius only kinetic descriptions provide useful tools for investigations.

Using a more detailed description of CR transport is clearly worthwhile from a theoretical standpoint but whether this is also the case for numerical simulations of marco-scale astrophysical environments is debatable. Modern day simulations are not only limited by the computational power available but also by the amount of physical processes modelled within them. Focusing more on micro-scale kinetic CR physical processes and how these can be coarse-grained and incorporated into a large-scale fluid model may lead to more correct solutions with higher physical credibility in comparison to higher-order CR fluid models. As CR transport is tightly linked to the magnetic field topology, using an accurate numerical method to calculate the evolution of the magnetic field will have a positive impact on the accuracy of the simulated CRs. Note that we can only observe minor differences in the simulations with CR scattering because of our high-resolution idealised setup. In resolution-starved simulations of realistic astrophysical scenarios, motions of the plasma and the magnetic field introduce another source of scattering via numerical diffusion. This additional but artificial scattering will further decrease the differences between the considered hydrodynamical models.

In summary, we find that the P1 approximation produces equally accurate solutions in comparison to the M1 approximation of CR transport {\em if CR scattering is efficient} while there are small differences if scattering is inefficient. However, the applicability of the M1 approximation is questionable in the case of inefficient scattering and higher-order fluid models such as the presented H1 model or fluid-kinetic models provide a better approximation, allow for more degrees of freedom to be evolved, and retain more information of the underlying CR distribution.

\section*{Acknowledgements}

TT and CP acknowledge support by the European Research Council under ERC-CoG grant CRAGSMAN-646955.

\section*{Data Availability}
The data underlying this article will be shared on reasonable request to the corresponding author.



\bibliographystyle{mnras}
\bibliography{main}


\appendix

\bsp	
\label{lastpage}
\end{document}